\documentstyle[prl,twocolumn,aps,epsf]{revtex}

\renewcommand{\narrowtext}{\begin{multicols}{2}\global\columnwidth20.5pc}
\renewcommand{\widetext}{\end{multicols}\global\columnwidth42.5pc}
\input{epsf.tex}

\begin{document}
\title{Spectral function of the electron in a superconducting RVB state}
\author{V.N. Muthukumar,$^a$ Z.Y. Weng,$^{b,c}$ D.N. Sheng$^d$}
\address{$^a$ Department of Physics, Princeton University, Princeton,
NJ 08544\\
$^b$ Center for Advanced Study, Tsinghua University, Beijing 100084\\
$^c$ Texas Center for Superconductivity, University of Houston, Houston, TX
77204\\
$^d$ Department of Physics and Astronomy, California State University,\\
Northridge, CA 91330}
\maketitle

\begin{abstract}
We present a model calculation of the spectral function of an electron in a
superconducting resonating valence bond (RVB) state. The RVB state,
described by the phase-string mean field theory, is characterized by three
important features: (i) spin-charge separation, (ii) short range
antiferromagnetic correlations, and (iii) holon condensation. The results of
our calculation are in good agreement with data obtained from Angle Resolved
Photoemission Spectroscopy (ARPES) in superconducting
Bi$_2$Sr$_2$CaCu$_2$O$
_{8+\delta}$ at optimal doping concentration.
\end{abstract}


The spectral function of an electron, $A({\bf k},\omega)$, being the
probability of finding an electron with momentum ${\bf k}$ and energy
$\omega
$, is a fundamental quantity in any description of interacting electrons.
ARPES is a very powerful and direct experimental technique used to measure
the spectral function in electronic systems. In the last decade, ARPES
measurements have contributed much to our understanding of the high
temperature superconductors (HTSC). Measurements done in the normal and the
superconducting state of the HTSC reveal a variety of interesting features
\cite{campuzano_aspen}, \cite{tohyama_00}. For instance, the normal state
spectra are extremely broad, with the broad peak evolving into a hump at
$T_c
$. In addition to the hump, a very sharp peak appears below $T_c$ at low
binding energies. This is observed most clearly near the Brillouin zone
boundary around the $M$ points, ($0,\pm\pi$) and ($\pm\pi,0$) \cite
{fedorov_99}. The strength of the sharp peak appearing below $T_c$ is
proportional to the superfluid density \cite{feng_00}, \cite{ding_00}. Along
the direction $(0,0) \rightarrow (\pi,\pi)$, there is some debate if the
data below $T_c$ can be interpreted as showing a clear break between a
coherent quasiparticle part and the broad incoherent background \cite
{valla_99}, \cite{campuzano_aspen}.

It was first noted by Anderson that spin-charge separation may provide a
natural explanation for the ARPES results \cite{pwa_book}. The basic idea is
that the hole created by a photon decays into a spinless holon and a neutral
spinon excitation. Such a decomposition can explain the broad incoherent
background seen in photoemission spectra, as well as the absence of a sharp
quasiparticle peak in the normal state. The emergence of a sharp
quasiparticle state below $T_c$ is attributed to the condensation of
holons. However, as we shall demonstrate in this paper, the inclusion of
short range antiferromagnetic (AF) correlations induced by the photohole is
crucial to the understanding of the ARPES results. This has been pointed out
in the context of photoemission from a Mott insulator \cite{zy_01}. In this
Letter, we consider the photoemission spectra observed in the
superconducting state of optimally doped Bi$_2$Sr$_2$CaCu$_2$O$_{8+\delta }$
(Bi 2212). We present a model calculation of the spectral function in a
superconducting RVB state, incorporating three important features: (i) short
range AF correlations, (ii) spin-charge separation, and (iii) holon
condensation. We show that such a description provides a consistent
explanation for many of the generic features observed in the superconducting
state of optimally doped Bi 2212.

{\em AF oscillations:} We begin with the decomposition of the electron
operator in the so-called phase string formulation, $c_{i\sigma}=h^{
\dagger}_i a_{i\sigma}$, where $h^{\dagger}_i$ is the bosonic holon creation
operator, and $a_{i\sigma}$ is a {\em composite} spinon operator that
satisfies fermionic anticommutation relations \cite{zy_97}. Unlike the
conventional slave-boson formalism, the (composite) spinon operator $a$ is
defined by $a_{i\sigma}\equiv b_{i\sigma} e^{i\hat{\Theta}_{i\sigma}}$,
where $b_{i\sigma}$ is the bosonic {\it elementary} spinon annihilation
operator and $\hat{\Theta}_{i\sigma}$, a nonlocal phase string operator
whose origin lies in the fact that a hole moving through a locally AF
background always picks up a string of sequential $\pm$ signs \cite{zy_97}.
The factor $\hat{\Theta}_{i\sigma}$ imposes anticommutation rules on $
a_{i\sigma}$ as well as $c_{i\sigma}$.

In the superconducting phase, the holons ($h_i^{\dagger }$) condense. An
elementary nodal (d-wave) fermionic excitation emerges in the spinon sector
and its creation operator $\gamma _{{\bf k}\sigma }^{\dagger }$ is related
to the $a$-operator, to leading order as \cite{zy_00} $a_{{\bf k}\sigma
}|\Psi _G\rangle \propto -\sigma v_{{\bf k}}\gamma _{{\bf k}\sigma
}^{\dagger }|\Psi _G\rangle $, where $v_{{\bf k}}$ is defined in the usual
manner as $v_{{\bf k}}={\rm sgn}(\Delta _{{\bf k}})(1-\xi _{{\bf k}}/E_{{\bf
k}})^{1/2}/\sqrt{2}$, with $E_{{\bf k}}^s=\sqrt{\xi _{{\bf k}}^2+\Delta _{
{\bf k}}^2}$ and $\Delta _{{\bf k}}=\Delta _0(\cos k_xa-\cos k_ya)$. Thus,
the superconducting ground state, $|\Psi _G\rangle $, in the phase string
description looks quite similar to the d-wave mean field state in the
slave-boson approach \cite{slaveboson}. However, the above
considerations
(obtained by using equations of motion), miss the local AF structure
around a hole that is created. To see this, let us consider a bare hole
state created by the $c$-operator, $c_{i\uparrow }|\Psi _G\rangle $ and
measure the spin configuration around the hole by evaluating the quantity $
\langle \Psi _G|c_{i\uparrow }^{\dagger }S_j^zc_{i\uparrow }|\Psi _G\rangle
.
$ Now,
\begin{eqnarray}\label{c1}
\langle \Psi _G|c_{i\uparrow }^{\dagger }S_j^zc_{i\uparrow }|\Psi _G\rangle
&=&\langle \Psi _G|b_{i\uparrow }^{\dagger }\frac 12\sum_\alpha \alpha
b_{j\alpha }^{\dagger }b_{j\alpha }b_{i\uparrow }|\Psi _G\rangle   \nonumber
\label{c1} \\
=-\frac 12|\langle \Psi _G|b_{i\uparrow }^{\dagger }b_{j\downarrow
}^{\dagger }|\Psi _G\rangle |^2 &+&\frac 12|\langle \Psi _G|b_{i\uparrow
}^{\dagger }b_{j\uparrow }|\Psi _G\rangle |^2~.
\end{eqnarray}
Using the mean field solution \cite{zy_00}, we get
\begin{equation}
\langle \Psi _G|b_{i\uparrow }^{\dagger }b_{j\downarrow }^{\dagger }|\Psi
_G\rangle =\left[ (-1)^{i-j}-1\right] \sum_mu_mv_mw_{m\uparrow
}^{*}(i)w_{m\uparrow }(j)~,  \label{t1}
\end{equation}
and
\begin{equation}
\langle \Psi _G|b_{i\uparrow }^{\dagger }b_{j\uparrow }|\Psi _G\rangle
=\left[ (-1)^{i-j}+1\right] \sum_mv_m^2w_{m\uparrow }^{*}(i)w_{m\uparrow
}(j)~.  \label{t2}
\end{equation}
In the above, $u_m$ and $v_m$ are the ``coherence factors'' of the mean
field theory determined self consistently for a given hole concentration $
\delta $, within a Bogoliubov-de Gennes scheme, and $w_{m\sigma }$, a one
particle wave function. Equations (\ref{t1}) and (\ref{t2}) are used in
(\ref {c1}) to determine the spin density around the hole site.
The results are
shown in Fig. 1, where we have plotted the spin density as a function of the
hole distance along the $\hat{x}$ axis for hole concentrations
$\delta =0$, and $\delta =1/7\simeq 0.14$.
As seen in the figure, the hole is surrounded by AF
oscillations in the local spin density. The oscillation is to be expected
from (\ref{c1}) and is indicative of local AF correlations.
The ground state $|\Psi _G\rangle $ is a singlet and the presence of
$c_{i\uparrow }|\Psi _G\rangle $
(an up spin hole) is accompanied by a spinon excitation with
$S^z=-1/2$. Assuming the (up spin) hole to be located on an
odd sublattice
site, our results show that the (down) spinon is created only on the even
sublattice sites. Note the presence of nonvanishing (up) spin density on the
odd sublattice sites, which we interpret as overscreening. Thus, the
combination of the up spin hole on an odd site and the down (up) spin
response on the odd (even) sites represents a spin-polaron with zero net
spin.

To account for this effect in determining the spectral function, we adopt
the following empirical approach. We write
\begin{equation}
a_{i\uparrow }|\Psi _G\rangle \simeq \sum_j\eta _j(i)\alpha _{j\downarrow
}^{\dagger }|\Psi _G\rangle +\mbox{higher order terms}~,  \label{spinonexp1}
\end{equation}
where $\eta _j(i)\neq 0$ only for $i$ and $j$ not belonging to the same
sublattice, and $\alpha _{{\bf k}\sigma }\equiv |v_{{\bf k}}|\gamma _{{\bf
k}
\sigma }$ (the d-wave sign in $v_{{\bf k}}$ will be absorbed into $\eta $).
We only retain the nearest and third nearest neighboring sites in the
expansion (\ref{spinonexp1}); {\em viz.}, $\eta _j(i)=+(-)\eta _0/2$, if $
j=i\pm \hat{x}(\hat{y})$; $\eta _j(i)=+(-)\eta _1/2$, if $j=i\pm 2\hat{x}(
\hat{x})\pm \hat{y}(2\hat{y})$; $\eta _j(i)\approx 0$ for all other sites.
The sign of $\eta _j(i)$ is from the d-wave symmetry of spinon pairing.
Transforming to momentum space, we get, within this approximation,
$a_{{\bf k}\uparrow }|\Psi _G\rangle \approx \eta _{{\bf k}}|v_{{\bf k}%
}|\gamma _{-{\bf k}\downarrow }^{\dagger }|\Psi _G\rangle ~,$
where
\begin{eqnarray}
\eta _{{\bf k}} &=&\eta _0(\cos k_xa-\cos k_ya)  \nonumber \\
&+&2\eta _1(\cos k_xa~\cos 2k_ya-\cos k_ya~\cos 2k_xa)~.
\end{eqnarray}
Here we take $\eta _1/\eta _0\simeq 0.3$
for $\delta =0.14$.

We may now ask what happens when an electron is created (as, for example, in
inverse photoemission), {\em i.e.}, $c^{\dagger}_{i\uparrow}
|\Psi_G\rangle$
. It can be shown in this case that the $\uparrow$ spinon is created at the
site $i$ with some residual amplitude extended over other sites of the same
sublattice. Neglecting the residual amplitude, to leading order we get
\[
a^{\dagger}_{{\bf k}\uparrow}|\Psi_G\rangle \simeq u_{{\bf k}
}\gamma^{\dagger}_{{\bf k} \uparrow}|\Psi_G\rangle~,
\]
which is essentially the same as the slave-boson
result. In the above, $u^2_{k}=1-v^2_{k}$.
Therefore, the main distinction between the usual d-wave slave-boson
mean field theory and this calculation is the momentum dependent factor $
\eta_{{\bf k}}$ that arises in the hole channel (corresponding to the
creation of a photohole in ARPES). The remainder of this paper is devoted to
a study of this channel. The differences between particle and hole
spectroscopies as well as a detailed comparison with slave-boson theories
will be presented elsewhere. 

{\em Spin-charge separation and holon condensation:} In the superconducting
state, the spinons are paired with $d$-wave symmetry as discussed earlier.
The effective hamiltonian for the holons is given by $H_h=-t_h\sum_{\langle
ij\rangle }e^{iA_{ij}^f}h_i^{\dagger }h_j+{\rm h.c.}$, where $A_{ij}^f$
represents the gauge field due to the spinons seen by the holons. Since the
spinons are paired, the mean field solution leads to the result $\sum_{\Box
}A_{ij}^f\approx -\pi $ \cite{zy_00}. Thus the superconducting state is
described by paired spinons and a Bose condensate of holons that experience
a $\pi $ flux around an elementary plaquette. Based on our earlier
discussion, we obtain the spectral function of the spinons, $\rho _a({\bf
k} ,\omega )=\eta _{{\bf k}}^2v_{{\bf k}}^2\delta
(\omega +E_{{\bf k}}^s)~.$
The spectral function for the holons is easily derived as
\[
\rho _h({\bf k},\omega )=\cos ^2{\frac{\theta _{{\bf k}}}2}\delta (\omega
-\epsilon _{{\bf k}-}^h)+\sin ^2\frac{\theta _{{\bf k}}}2\delta (\omega
-\epsilon _{{\bf k}+}^h)~,
\]
where $\cos \theta _{{\bf k}}=\gamma _{{\bf k}}/(\sqrt{2}\lambda _{{\bf
k}})$. Here, $\gamma _{{\bf k}}=\cos k_xa+\cos k_ya$,
$\lambda _{{\bf k}}=\sqrt{
\cos ^2k_xa+\cos ^2k_ya}$, and
$\epsilon _{{\bf k}\pm }^h=\pm 2t_h\lambda
_{ {\bf k}}-\mu _h$. At $T=0$, the holon chemical potential
$\mu _h=-2t_h\lambda _{{\bf k}=0}=-2\sqrt{2}t_h$.

We now construct the spectral function of the electron by using
the operator decomposition of the electron,
$c_{i\sigma}=h_0a_{i\sigma}+c^\prime_{i\sigma}
$, where $<h^\dagger_i>=h_0$, describes the Bose condensate of holons, and
$
c^\prime_{i\sigma}=:h^\dagger_i:a_{i\sigma}$, with $:h^\dagger_i: \equiv
h^\dagger_i-h_0$. The spectral function of the electron $A^{{\rm e}}_-({\bf
k
},\omega)$, is expressed as a convolution of the spinon and holon spectral
functions. It is easy to see that
\begin{equation}
A^{{\rm e}}_{-}({\bf k},\omega)=\theta(-\omega)\frac 1 N \sum_{{\bf k}}
\int^0_{\omega} d\omega^{\prime}\rho_h({\bf k}^{\prime}-{\bf k}
,\omega^{\prime}-\omega) \rho_a({\bf k}^{\prime},\omega)~,
\end{equation}
where $\rho_a$ and $\rho_h$ are the spectral functions of the spinon and
holon respectively. For obvious reasons, we write the spectral function as
the sum of an incoherent and a coherent part, $A^{{\rm e}}_-= A^{{\rm i}}_-
+A^{{\rm c}}_-$. Clearly, these two terms correspond to the two terms in the
electron decomposition.

The coherent part of the spectral function is obtained from the contribution
of the holon condensate. We get
\begin{equation}  \label{coherent}
A^{{\rm c}}_-({\bf k},\omega)=\rho_h^c\eta_{{\bf k}}^2v^2_{{\bf k}
}\delta(\omega+E^s_{{\bf k}})~,
\end{equation}
where $\rho_h^c \propto \delta$ denotes the density of the holon condensate.
This contribution is dubbed ``coherent'', since it is a sharp peak appearing
below $T_c$ (the temperature at which the holons condense). The incoherent
part of the spectral function is the convolution,
\begin{eqnarray}  \label{incoherent}
A^{{\rm i}}_-({\bf k},\omega)&=& {\frac{1 }{N}} \sum_{{\bf k}
^\prime}{}^\prime \eta^2_{{\bf k}^\prime+{\bf k}}v^2_{{\bf k}^\prime+{\bf
k}
} [\cos^2 {\frac{\theta_{{\bf k}}}{2}} \delta(\omega+E^s_{{\bf k}^{\prime}+
{\bf k}}+\epsilon^h_{{\bf k}^{\prime}-})  \nonumber \\
& & \mbox{} +\sin ^2 {\frac{\theta_{{\bf k}}}{2}} \delta(\omega+E^s_{{\bf
k}
^{\prime}+{\bf k}}+\epsilon^h_{{\bf k}^{\prime}+})]~.
\end{eqnarray}
The prime in the summation indicates that the contribution from the holon
condensate is removed.

{\em Comparison with ARPES results:} We are now in a position to plot $
A_{-}^{{\rm e}}({\bf k},\omega )$ for various momenta and compare with
results from ARPES. For the calculation of the spectral function, we choose
the following parameters. The dispersion of the spinons, $\xi _{{\bf k}}$,
is determined by the nearest and next nearest neighbor hopping integrals, $
t_1=75$ meV, $t_2=20$ meV. The chemical potential, $\mu =-62$ meV is chosen
to mimic the topography of the observed Fermi surface. We choose a value of
$
\Delta =20$ meV for the gap, and $t_h=4t_1=0.3$ eV for the holon dispersion.
Before proceeding to discuss the results for $A_{-}^{{\rm e}}({\bf k},\omega
)$, we note that the following can be anticipated: (a) The coherent part of
the spectral function, $A_{-}^{{\rm c}}({\bf k},\omega )\propto h_0^2\propto
\delta $. So, the strength of the coherent peak will be proportional to the
superfluid density, as observed experimentally \cite{ding_00},
\cite{feng_00}. (b) As is evident from our results, $A_{-}^{{\rm c}}({\bf
k},\omega )$ is
{\em strongly} momentum dependent, owing to the factor $\eta _{{\bf k}}^2$.
This contribution is strongest around the $M$ point and weakest around the
$\Gamma $ point. (c) The incoherent part of the spectral function is
expected
to produce a broad background. Such a broad background is always seen in
photoemission. Since, in our calculation, the background arises from a
convolution of spinon and holon spectral functions, we expect a rather {\em
weak} momentum dependence of the broad background. This may have
already been observed experimentally \cite{valla_unpubl}.

In Fig. 2, we show the results for $A_{-}^{{\rm e}}({\bf k},\omega )$ at a
point on the Fermi surface, ${\bf k}=(0.4,\pi )$. We see clearly that the
total spectral response is the sum of the incoherent and coherent pieces of
the spectral function. The sharp peak at lower binding energy corresponds to
$A^{{\rm c}}$ and the peak is located at an energy $\omega =E_{{\bf k}}^s$.
The broad feature seen in the figure is the contribution from $A^{{\rm i}}$.
The incoherent part exhibits a
low energy edge around $E_{{\bf k}
}^s$, and a broad peak (``hump'') slightly above it. The origin of these
features is very much the same as in the spectra of the undoped insulator
\cite{zy_01}. In both cases, the low energy edge is determined by the
dispersion of the spinon, $E_{{\bf k}}^s$. The origin of the hump lies in
the local AF correlations embodied in the factor $\eta _{{\bf k}}$. Let us
consider (\ref{incoherent}). The factor $\eta _{{\bf k}}$ is maximum at the
$ M$ point. As $\omega $ increases, ${\bf k}^{\prime }\neq 0$
terms in (\ref
{incoherent}) have to contribute to the sum. However, for such terms, the
factor $\eta _{{\bf k}+{\bf k}^{\prime }}$ decreases, thereby causing a hump
in the spectral funtion. The position of the hump is shifted from the edge
by $\epsilon _{{\bf k}^{\prime }-}^h$. A similar effect occurs in the
photoemission of a single hole in the Mott insulator, where a broad
hump arises from the coherence factors of the Schwinger boson mean field
theory. We emphasize that the factor $\eta _{{\bf k}}$ in the present case
and the coherence factors in the Schwinger
boson theory of the insulator reflect the antiferromagnetic
correlations that play an important role in the formation of the observed
humps. Seen in this perspective,
the hump is not directly related to the 41 meV
resonance observed in the neutron scattering \cite{hong_95},
as conjectured in the literature \cite{campuzano_99} though AF
correlations clearly play a crucial role in both explanations.

In Fig. 3, we show the behavior of $A_{-}^{{\rm e}}({\bf k},\omega )$ for
various ${\bf k}$ points. In the left panel, we show how the spectral
function evolves as one moves away from the $M$ point in two
perpendicular directions. As can be seen in the figure, the peak-dip-hump
structure gets more pronounced near the $M$ point. In the right panel, we
show the spectral function for a series of points on the Fermi surface.
Again, we see that the factor $\eta _{{\bf k}}$ causes the coherent peak as
well as the hump to diminish away from the $M$ point. In particular, along
the direction $\Gamma \rightarrow (\pi ,\pi )$, these features are absent in
our calculation. Though this is an approximate result, it shows that the
inclusion of AF correlations suppresses the peak-dip-hump structure along
this direction.

To conclude, our model calculation of the spectral function of an electron
in a superconducting RVB state explains many of the generic features
observed
in the photoemission spectra of optimally doped Bi 2212. We find that the
inclusion of short range antiferromagnetic correlations is extremely crucial
to the understanding of these features. In particular, we show that the
presence of a hole induces Friedel-like oscillations in the local spin
density surrounding the hole. We propose a scheme to incorporate this effect
in the calculation of the spectral function. When this is taken into account
in conjunction with spin-charge separation and holon condensation below
superconducting $T_c$, many of the features seen in photoemission can be
explained naturally. Our calculation shows that these features have
analogues in the photoemission spectra of the undoped Mott insulator where
the short range antiferromagnetic correlations play a very important role
too. \noindent

We thank P. W. Anderson for his comments. We are also thankful to H. Ding,
A. Fedorov, P. D. Johnson, T. Valla and B. O. Wells for several discussions
on ARPES. V.N.M. is supported by NSF Grant DMR-9104873. V.N.M. also thanks
the hospitality at Texas Center for Superconductivity, University of Houston
where part of this work was carried out.

{Fig. 1 Spin configuration near an up-spin hole, created by $c_{\uparrow}$
on the
ground state at $\delta=0$ and $0.14$, along the $\hat{x}$
axis. }\\

{Fig. 2 The spectral function at ${\bf k}=$($0.4$, $\pi$), which is a Fermi
point at the Brillouin zone boundary (see the inset).
The dashed curve shows only the incoherent part whose low energy edge
coincides
with the coherent peak position while its `hump' lies slightly above the
edge.}\\

{Fig.3 The spectral function at different momenta marked by full circles
in the insets of two panels. The left panel corresponds to scans near
the M point along two directions, while the right represents momenta
on the Fermi surface.}
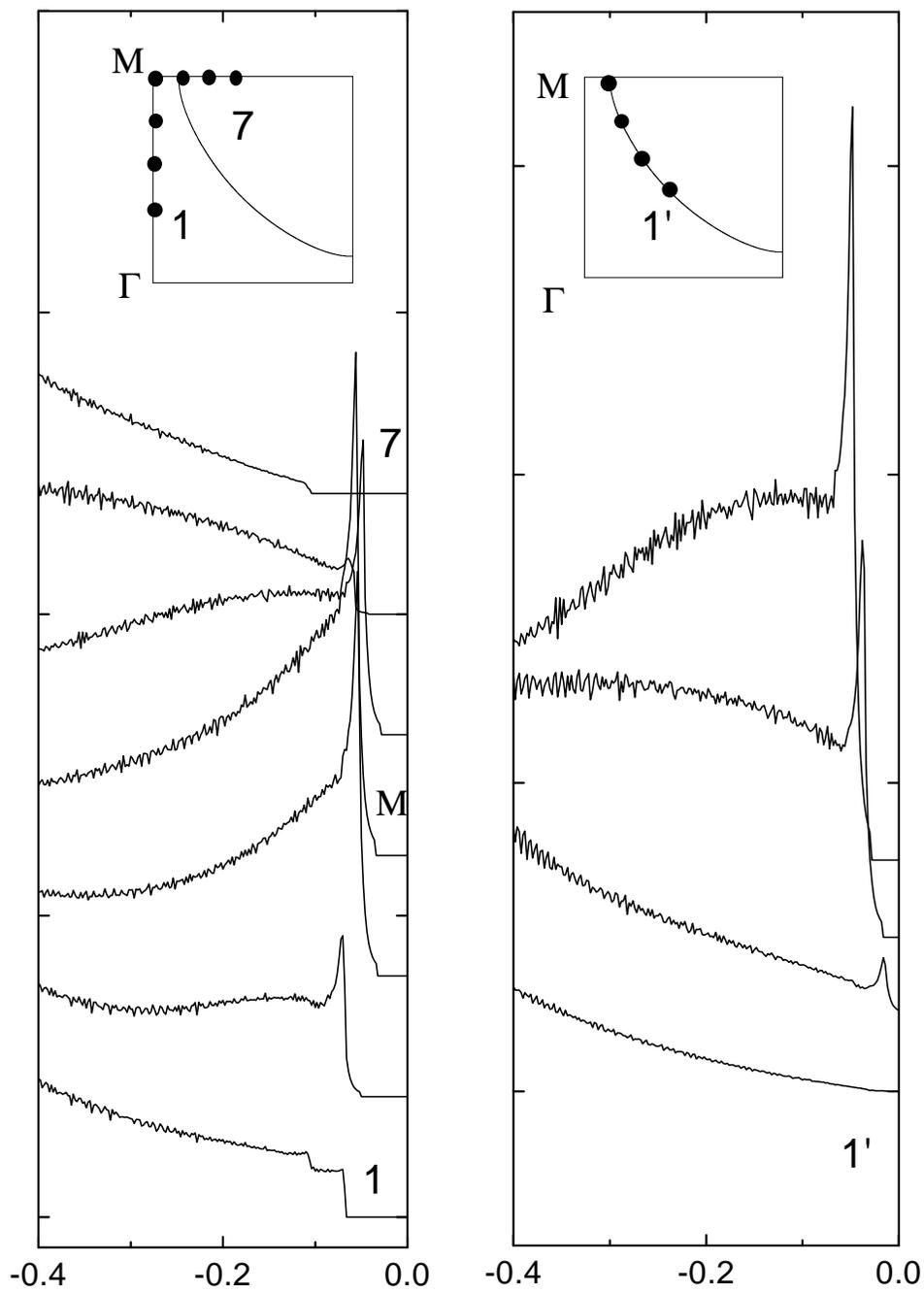
\end{document}